\input psfig.sty
\documentclass{kapproc} 
\kluwerbib

\begin{document}
\articletitle{Recent star formation in very~ \\ 
 luminous infrared galaxies}
\author{Alessandro Bressan, Bianca Poggianti}
\affil{Osservatorio Astronomico di Padova, Padova,IT}
\author{Alberto Franceschini}
\affil{Dipartimento di Astronomia, Universita' di Padova, Padova,IT}
\begin{abstract}
Among star forming galaxies, a spectral combination of
a strong 
$\rm H\delta$ line in absorption and a moderate [O{\sc ii}]
emission has been suggested
to be a useful method to identify dusty starburst galaxies at any redshift,
on the basis of optical data alone.
On one side it has been shown that such a spectral particularity is
indeed suggestive of obscured starburst galaxies but, on the other, 
the degeneracy of the optical spectrum has hindered any quantitative
estimate of the star formation and extinction during the burst.
The optical spectrum by
itself, even complemented with the information on the far-IR flux, 
is not enough to
identify univocal evolutionary patterns. 
We discuss in the  following whether 
it is possible to  reduce these uncertainties by
extending the spectral analyses to the near infrared
spectral region.
\end{abstract}


\section{Introduction}
Recent observations in the IR/mm have revealed that luminous and ultra luminous
IR galaxies (LIRGs and ULIRGs) constitute a major cosmological component during 
the past epochs of the universe:
in spite of their short duration, the transient IR-active phases are
responsible for a large fraction
of the energy emission and metal production at high redshifts. 
With luminosities spanning the range  10$^{11}$--10$^{13}$ L$_{\odot}$
and space densities similar to those of quasars (Soifer et al. 1986)
Luminous and Ultra luminous Infrared galaxies (LIRGs and ULIRGs) are the most luminous
objects in the local Universe.  Highly extinguished and strong IR
emitters, ULIRGs are considered to be the local  analogues of the
newly discovered high-z IR luminous galaxies.  Recent optical
spectroscopic surveys (Wu et al. 1998, Poggianti and Wu 2000) reveal
that the spectra of a large fraction of 
(U)LIRGs display a peculiar combination of spectral features in the
optical: a strong $\rm H\delta$ line in absorption (EW$> 4$ \AA) and a moderate
[O{\sc ii}] emission (EW$< 40$ \AA). 
Galaxies with this type of spectra were named ``e(a)'' galaxies
and were found to be quite numerous in the cluster and
field environments at $z=0.4-0.5$ (Poggianti
et al. 1999). The equivalent width of their $\rm H\delta$ line
exceeds that of typical, quiescent spirals at low-z and their
low [O{\sc ii}]/$\rm H\alpha$ ratios are consistent with
the emission line fluxes being highly extincted by dust.
Until recently, such a combination of
moderate [OII] emission and strong Balmer absorption, was thought to
be associated with  post-starburst galaxies, with little or no
star-formation. Such a claim was mostly based on dust-free models.
Poggianti et al. 1999 however, proposed that this peculiar spectral
signature corresponds in fact to highly obscured starburst galaxies.
Clearly, one still needs to investigate the exact origin of these
spectra,  namely, the nature of the star formation as well as the
properties of the dust.  A possible explanation for this unusual
combination of emission$/$absorption features is selective extinction
(Poggianti and Wu 2000). In such a scenario,  HII regions (where the
[OII] emission originates) are highly embedded and thus are affected
by more extinction compared to the older stellar population which is
responsible for the Balmer absorption.

Surprisingly enough, the combination of moderate emission lines and strong 
Balmer absorption has been also detected in the spectra of high-z IR luminous ISO
galaxies (e.g. Flores et al. 1999). 
A recent near-IR 
survey of the same population revealed that the ISO galaxies are in fact
actively starbursting but highly obscured objects, based on the strength of the 
H$_{\alpha}$ emission line and high equivalent width (Rigopoulou et al. 2000).
Differential extinction is again at the heart of this spectral behaviour.
Rigopoulou et al. (2000) have estimated the extinction in their sample of
ISO galaxies based on optical colours and found it to be A$_{v} \sim$3.
After correcting the H$_{\alpha}$ line flux using optical indicators of
extinction, the inferred Star Formation Rates (SFR) 
fall significantly below the proper SFR estimated from the far-Infrared flux.
This implies that optical extinction is significant, and requires additional
information for a more reliable evaluation.

Poggianti, Bressan and Franceschini (2001) have recently investigated in some detail the
optical spectra of LIRG and ULIRG galaxies, to constrain the
recent history of star-formation and the dust extinction characteristic of various
stellar populations. By examining different star formation patterns
they concluded that only a starburst, selective extinction scenario
could explain the e(a) spectra. In this scenario the extinction is larger in the younger
populations, a fact that mimics the progressive escape of young stars from their
parental molecular clouds, as they age. The observed FIR/V luminosity ratio could be
explained only by models where a significant fraction of the FIR luminosity originates
in regions that are practically obscured at optical wavelengths. Unfortunately this
hinders any quantitative estimate of the duration and intensity of the burst and/or of
the wavelength dependence of the extinction. Thus the optical-near UV spectrum by
itself, even complemented with the information on the far-IR flux, are not enough to
identify univocal modellistic solutions. One chance to reduce these uncertainties is
provided by the extension  of the spectral analyses to the near-IR regime 
(eg. Murphy et al. 1999).

\section{Models in the near IR}

In order to highlight the advantages of investigating over a 
wider wavelength range, we have followed the same methodology
described in Poggianti, Bressan \& Franceschini (2001).
The integrated model spectrum, from the far UV to the near IR,
has been generated as a combination 
of 10 stellar populations of different ages, computed with a Salpeter IMF 
between 0.15 and 120 $M_{\odot}$.
The stellar SEDs have been obtained by extending
the Pickles spectral library below 1000 \AA $\,$ and above
24000 \AA with the Kurucz (1993) models.
The composite (stars+gas) spectrum of each single generation 
has been then produced by one of us (AB) with the help of the
photo-ionization code CLOUDY (Ferland, 1990).
The ages of the 10 populations have been 
chosen considering the evolutionary time scales associated with the 
observational
constraints: the youngest generations ($10^6, 3 \cdot 10^6, 8 \cdot 10^6,
10^7$ yr) are responsible for the ionizing photons that produce the 
emission lines;
the intermediate populations ($5 \cdot 10^7, 10^8, 3 \cdot10^8, 5 
\cdot 10^8, 10^9$ yr) are those with the strongest Balmer lines in 
absorption, while older generations of stars 
have been modelled as a constant star formation rate 
(SFR) between 2 and 12 Gyr before the moment of the observation
and can give a significant contribution to the spectral continuum, hence
affecting also the equivalent widths of the lines.

Each simple population is assumed to be extincted by dust
in a uniform screen according to the standard extinction law of the 
diffuse medium in our Galaxy ($R_V=A_V/E(B-V)=3.1$, Cardelli et al. 1989). 
While a more complex picture of the extinction cannot be excluded,
Poggianti et al. already have shown that
the characteristics of the emerging
spectrum require a significant amount of {\sl foreground} dust 
(screen model). 
Indeed in the case of a uniform mixture of
dust and stars, increasing the obscuration does not yield a corresponding
increase in the {\em reddening} of the spectrum: the latter saturates
to a value (E(B-V)$\sim 0.18)$) which is
too low to be able to account for the observed emission line
ratios (see also Calzetti et al. 1994).
In the model used here, the extinction value E(B-V)
is allowed to vary from one stellar population to another and 
the extincted spectral energy distributions of all the single generations
are added up to give the total integrated spectrum.

Finally, the best-fit model, within a chosen star formation scenario, 
was obtained by
minimizing the differences between selected features in the observed
and model optical range: the equivalent width of four lines
([OII]$\lambda$3727,H$\delta$, H$\beta$ and H$\alpha$) and the
relative intensities of the continuum flux in eight almost featureless
spectral windows (3770-3900\AA, 4020-4070\AA, 4150-4250\AA,
4600-4800\AA, 5060-5150\AA, 5400-5850\AA, 5950-6250\AA $\,$ and
6370-6460\AA).

\begin{figure}[h]
\psfig{file=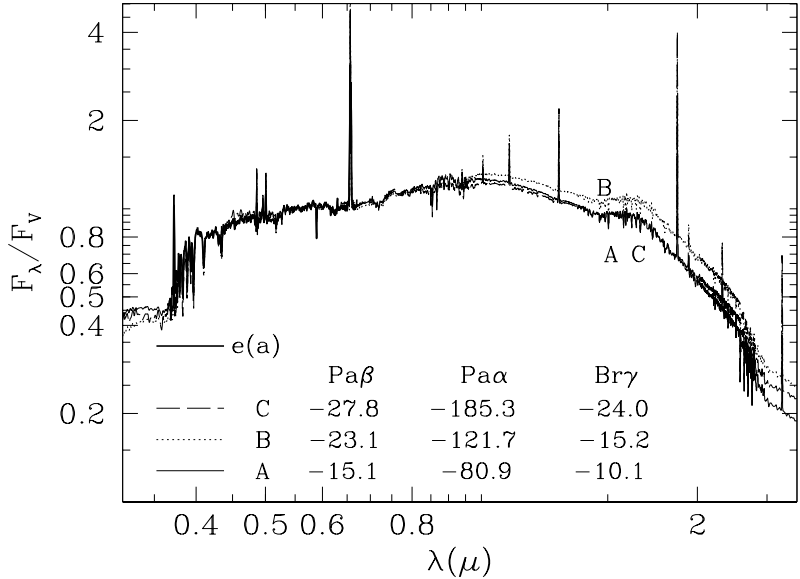,angle=270,width=12.5cm}
\caption{The observed average e(a) optical (3700\AA -- 6900\AA) spectrum is
compared with suitable models extended into the NIR
region. All models reproduce well the
shape of the observed continuum and equivalent widths of
H$\alpha$ (-62.48$\pm$5.3), H$\beta$ (0.68$\pm$0.8), 
H$\delta$ (5.64$\pm$0.5) and [OII]3727 (-12.68$\pm$2.1)(Poggianti et al
2001). The predicted EWs of selected lines in the NIR are shown. See text for more details.}
\label{fig1}
\end{figure}

\section{Results}

Fig.\ref{fig1} depicts three particular cases that were
successful in reproducing the optical spectral features of e(a) galaxies
(Poggianti et al. 2001).
Model A, the reference model, reproduces both the line strengths and
the highly reddened continuum, but it can account for only 1/3 of the
FIR emission, hence of the star formation rate. In this model
the mass of young stars amounts to about 10\% of the galaxy mass.

To solve the above discrepancy, some practically obscured regions were
added in model B. As in model A, the starburst began $2 \cdot 10^8$ yr
ago, but both the SFR during the burst and the extinction of the two
youngest stellar generations are higher than in A. The recent SFR is
about a factor of 10 larger than in model A and, for a standard
Salpeter IMF in the range $0.1-100 \, M_{\odot}$, about 60\% of the
total mass in stars is formed during the burst. Thus the high ratio
FIR/optical in this case, is indicative of a significant fraction of
the galaxy mass being formed during the burst. It is worth noticing
that assuming a top-heavy IMF during the burst phase would
substantially reduce the mass fraction in young stars: for example,
with a Salpeter IMF lower mass limit $=1 \, M_{\odot}$, the starburst
in model B forms about 40\% of the total stellar mass. However
different IMF with the same distribution of stars above say
1$M_{\odot}$, would essentially produce the same integrated spectrum,
and would be practically indistinguishable.
Finally, in model C the observed FIR/optical ratio is reproduced by
adopting an extinction law with $R_V=A_V/E(B-V)=5$, as observed
towards some dense clouds in our Galaxy (Mathis 1990). In this case
the recent SFR is only a factor of two larger than that of model A and the 
observed FIR/optical ratio is fully reproduced because, for a given E(B-V), the
optical flux is much more extinguished.
In spite of the different star formation histories, the shape of the
continuum in the near infrared is fairly similar in the three cases.
This is striking because the inspection of the models shows that,
while in case A the near infrared continuum is 
contributed
in almost equal parts by the young burst and the old disk, in case B
it is  dominated by the young burst while, in case C, it is due to the
old disk.

Thus it
appears that the degeneracy encountered at optical wavelengths would
not be removed by the NIR photometry alone.

On the other hand the simulations clearly
suggest different equivalent widths of the NIR hydrogen lines.
In particular, in models B and C the line intensities are about twice
as strong as in model A. In the case of model B,
though the SFR is much  higher than
in the case A, the young
populations are much more heavily obscured so that line intensities 
which are similar in the optical, are only a factor of two larger in the NIR.
A significant fraction of the SFR remains hidden even in the NIR.
In the case of model C the intensity of the NIR lines simply reflects
the larger SFR adopted. 
\begin{table}[ht]
\centering
\caption{ E(B-V) from Hydrogen line ratios in the different models.}
\begin{tabular}{lccccc}
\hline
 & H$\beta$/H$\alpha$ & Pa$\alpha$/Br$\gamma$  &
  Pa$\beta$/Br$\gamma$  & H$\beta$/Br$\delta$  &
 H$\alpha$/Pa$\alpha$ \\ 
 A &      0.96 &  0.20 &  1.15 &  1.08 &  1.19 \\
 B &      1.19 &  0.72 &  1.45 &  1.35 &  1.47 \\
 C &      0.90 &  1.37 &  1.98 &  1.38 &  1.67 \\
 C$^a$ &  0.71 &  0.70 &  1.01 &  0.90 &  0.98 \\
\hline
\multispan{6}{\small $^a$ adopting an extinction law with R$_V$=5 \hfill} 
\end{tabular}
\label{ebv}
\end{table}
Finally, 
Table \ref{ebv} shows the average E(B-V) recovered with the 
ratios of the Hydrogen lines, in the optical and NIR spectral region,
adopting the extinction law with R$_V$=3.1.
We remind that, in the case of VLIRGs where the EW H$\beta$ is almost null, the 
Balmer decrement is severely affected by the assumed 
equivalent width of the absorption component.
On the other hand the
NIR ratio Pa$\alpha$/Br$\gamma$ is affected by too small a
wavelength range. As for the other three line ratios,
the Pa$\beta$/Br$\gamma$ ought to be preferred because
it is less affected by uncertainties in the old population
and unaffected by [NII] emission.
\begin{acknowledgments}
We are grateful to Pasquale Panuzzo
for his help in running CLOUDY and for discussions.
AB acknowledges support from the TMR grant ERBFMRXCT960086.     
\end{acknowledgments} 
{}
\smallskip
\end{document}